\documentclass[prl,aps,twocolumn,superscriptaddress,showpacs]{revtex4}
\usepackage{amsmath}
\usepackage{amsbsy}
\usepackage{amssymb}
\usepackage[dvips]{graphicx}
\usepackage{epsfig}
\usepackage{color}

\newcommand{\C}{\mathcal{C}}
\newcommand{\G}{\mathcal{G}}
\newcommand{\Z}{\mathcal{Z}}
\newcommand{\N}{\mathcal{N}}
\newcommand{\e}{\epsilon}

\renewcommand{\P}{\mathbb{P}}

\begin{document}

\title{Exactly solvable models of adaptive networks}

\author{Olivier Rivoire} 
\affiliation{Laboratory of Living Matter, The
Rockefeller University, 1230 York Ave., New York, NY-10021, USA.}

\author{Julien Barr\'e}
\affiliation{ Laboratoire J. A. Dieudonn\'e, Universit\'e de Nice-Sophia 
Antipolis, Parc Valrose, F-06108 Nice Cedex 02, France.}

\date{\today}

\begin{abstract}

A satisfiability (SAT-UNSAT) transition takes place for many
optimization problems when the number of constraints, graphically
represented by links between variables nodes, is brought above some
threshold. If the network of constraints is allowed to adapt by
redistributing its links, the SAT-UNSAT transition may be delayed and
preceded by an intermediate phase where the structure self-organizes
to satisfy the constraints. We present an analytic approach, based on
the recently introduced cavity method for large deviations, which
exactly describes the two phase transitions delimiting this adaptive
intermediate phase. We give explicit results for random bond models
subject to the connectivity or rigidity percolation transitions, and
compare them with numerical simulations.

\end{abstract}


\maketitle

Unraveling the principles responsible for the structure of observed
technological, sociological or biological networks is currently an
intensively pursued challenge~\cite{Newman03}. Phase transitions
occurring when the structure evolves are of particular interest. The
simplest example is connectivity percolation, which takes place when
the number of links increases~\cite{StaufferAharony92}. Analogous
phase transitions are found in constraint satisfaction problems
(CSP's) defined on random graphs, such as the $K$-SAT or coloring
problems~\cite{PapadimitriouSteiglitz82}. They are referred to as
SAT-UNSAT transitions and are related to the
algorithmic complexity of solving these hard combinatorial
problems~\cite{MonassonZecchina99}.
 
While past studies of SAT-UNSAT transitions have confined to random
distributions of constraints~\cite{MezardZecchina02,MuletPagnani02},
it has recently been suggested that an additional phase transition
could arise if the network is allowed to respond to the addition of
constraints by reorganizing
itself~\cite{ThorpeJacobs00,BarreBishop05b}. In this scenario, an
adaptive intermediate phase (AIP) is predicted where the system avoids
the SAT-UNSAT transition and adopts a structure distinct from random
graphs. In this letter, we consider models whose configuration space
consists of an ensemble of graphs, with a CSP defined on each graph, and we investigate the presence of an AIP by solving these models analytically with the cavity method for large deviations~\cite{Rivoire05}. We thus provide one of the very few analytical
results available for exponential random graphs models, the general
class of networks to which our models belong (see
e.g.~\cite{Newman03}).

We apply our general approach to rigidity percolation on random bond
models; this is a family of CSP's initially designed to model network
glasses, the physical system for which the presence of an AIP was
first suggested~\cite{ThorpeJacobs00} and experimentally
investigated~\cite{SelvanathanBresser00}. In the absence of
adaptation, these materials are expected to undergo a rigidity
transition when the mean coordination of their atoms is varied by
modifying their composition~\cite{Phillips79,Thorpe83}. As recalled
below, this rigidity percolation can be viewed as a particular example
of SAT-UNSAT transition~\cite{BarreBishop05a}. Random bond models
include models with a continuous connectivity percolation transition
or a discontinuous rigidity percolation transition, which allows us to
illustrate the necessity of a discontinuous SAT-UNSAT transition for
observing an AIP. For rigidity percolation, we
analytically describe this AIP and the related phase transitions by
deriving new and presumably exact formul\ae\ for large deviation
functions.

%%%%%%%%%%%%%%%%%%%%%%%%%%%%%%%%%%%%%%%%%%%%%%%%%%%%%%%%%%%%%%%%%%%%%%%%%%

\paragraph*{Random bond models ---}

Random bond models~\cite{DuxburyJacobs99} are constructed from $N$
point-like particles with $d$ degrees of freedom each, by adding
successively and at random bonds between them. Each bond carries only
a bond-stretching constraint, so that the addition of a bond can have
two effects: if the distance between its two end-points is already
constrained, the new bond does not modify the total number of degrees
of freedom and is considered as redundant; in the opposite case, it
suppresses the degree of freedom associated with the relative distance
between the two end-points. Globally, the number of independent
internal degrees of freedom, or floppy modes, can be written
\begin{equation}
\N_{\rm floppy}=dN-M+E_r-d(d+1)/2,
\end{equation}
where $dN$ represents the total number of degrees of freedom, $M$
the number of bonds, $E_r$ the number of redundant
constraints, and where $d(d+1)/2$ accounts for the global degrees
of freedom in $d$ dimensions.

In the thermodynamical limit where $N\to\infty$ and $M\to\infty$ with
fixed ratio $\alpha=M/N$, the structure is a random graph with Poisson
degree distribution~\cite{Bollobas01}, and the density of floppy
modes, $n_f\equiv\N_{\rm floppy}/N$, satisfies
\begin{equation}
n_f=d-\alpha+\e_r,
\end{equation}
with $\e_r\equiv E_r/N$. Two different phase transitions may occur
when $\alpha$ is increased: rigidity percolation, when the largest
rigid subgraph, with no floppy mode, becomes extensive; unstressed to stressed transition, when the largest stressed subgraph, with a non-zero density of redundant constraints, becomes extensive.  With random bonds, the two transitions occur in
fact simultaneously at a critical value
$\alpha_c$~\cite{ThorpeJacobs99}.

An elementary counting argument, due to Maxwell~\cite{Maxwell64},
approximatively locates $\alpha_c$ by balancing the total number of
constraints $M$ with the total number of degrees of freedom, $dN$,
yielding $\alpha_{\rm Mxl}=d$. The underlying assumption is that no
redundant bond appears ($\e_r=0$) before the number of floppy modes
vanishes ($n_f=0$), which is not the case with random networks where
the density of floppy modes never reaches zero (therefore
$\alpha_c<\alpha_{\rm Mxl}$).

%%%%%%%%%%%%%%%%%%%%%%%%%%%%%%%%%%%%%%%%%%%%%%%%%%%%%%%%%%%%%%%%%%%%%%%%%%

\paragraph*{Constraint satisfaction ---}

Random bond models can be recasted in the broader framework of CSP's as
follows~\cite{BarreBishop05a}. An instance of the problem is given by
a graph $G$. An arrow is assigned to each link of $G$, which must
point toward one of the two adjacent sites, and the set of different
orientations defines the set of admissible solutions. Such a
configuration $\sigma$ is labeled by associating to each oriented link
$i\to j$ a Boolean variable $\sigma_{i\to j}\in\{0,1\}$ with
$\sigma_{i\to j}=1$ if the arrow is directed from $i$ to $j$ (and
$\sigma_{j\to i}=1-\sigma_{i\to j}$). The cost function is
\begin{equation}
\C_G[\sigma]=\sum_i [ \max ( 0,d-\sum_{j\in i}\sigma_{j\to
i})-(d-\alpha)]
\end{equation}
where $j\in i$ indicates that $j$ is connected to $i$.  Physically, an
arrow means that the bond suppresses one degree of freedom to the node
to which it points. By minimizing the cost function, we get the number
of redundant constraints for the graph $G$~\cite{BarreBishop05a}:
\begin{equation}\label{eq:Er}
E_r[G]=\min_\sigma \C_G[\sigma].
\end{equation}
Note that $\C_G[\sigma]\geq 0$ for any $\sigma$, so that $E_r[G]\geq
0$ as it should. We will focus on two particular cases: $d=2$ which
describes 2$d$ rigidity percolation, and $d=1$, which describes the
usual connectivity percolation.

%%%%%%%%%%%%%%%%%%%%%%%%%%%%%%%%%%%%%%%%%%%%%%%%%%%%%%%%%%%%%%%%%%%%%%%%%%

\paragraph*{Infinite temperature transition ---}

\begin{figure}
\centering\epsfig{file=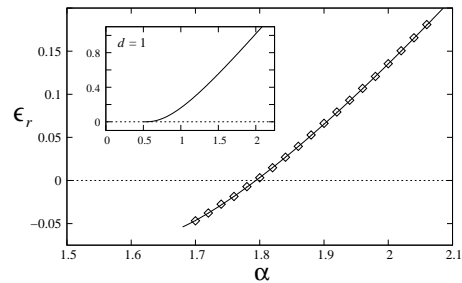,width=6cm} 
\caption{Density of redundant constraints $\e_r$ for $d=2$ as
calculated from Eqs.~\eqref{eq:er}; the symbols are from numerical
simulations confirming that the negative part corresponds to the number
of floppy modes on the 3-core. Inset: $\e_r$ for $d=1$.\label{fig:er}}
\end{figure}

The formulation in terms of CSP allows to resort to the cavity
method~\cite{MezardParisi03} to determine the phase diagram as a
function of the density of constraints $\alpha$, when $G$ is taken at
random from the ensemble $\G^{(\alpha)}_N$ of graphs with $N$ nodes
and $M=\alpha N$ links~\cite{BarreBishop05a}. This method indeed
applies to any CSP defined on such random graphs, and has notably lead
to the derivation of the phase diagrams of the $K$-SAT and coloring
problems~\cite{MezardZecchina02,MuletPagnani02}. In contrast with
these algorithmically hard CSP's, the present model can be solved by
the simplest, ``replica symmetric''~\cite{MezardParisi87b},
formulation of the cavity method. The density of redundant constraints
thus obtained is consistent with previous studies~\cite{Moukarzel03}
and reads
\begin{equation}\label{eq:er}
\e_r=\sum_{k=d+1}^\infty
\pi_{2\alpha\eta}(k)\left(\frac{k}{2}-d\right),\quad
\eta=\sum_{k=d}^\infty\pi_{2\alpha\eta}(k),
\end{equation}
where $\pi_{\theta}(k)=e^{-\theta}\theta^k/k!$.  A non-zero solution,
$\eta>0$, exists only for $\alpha>\alpha_p^{(d)}$. This threshold
corresponds to the percolation of the $(d+1)$-core~\cite{Moukarzel03},
that is, the emergence of a sub-graph containing an extensive number
of nodes all having connectivity at least $(d+1)$~\cite{Bollobas01}
(only the percolation of the 2-core occurs simultaneously with
connectivity percolation~\cite{Bollobas01}). As shown in
Fig.~\ref{fig:er}, for $d\geq 2$, when this non-zero solution appears,
the corresponding $\e_r$ is negative, and becomes positive only at
$\alpha_c^{(d)}>\alpha_p^{(d)}$.

Such a negative prediction for an intrinsically positive quantity is
often indicative of the inadequacy of the replica symmetric
assumption~\cite{MezardZecchina02,MuletPagnani02}. For the present
problem however, we verified that no replica symmetry breaking could
occur and found instead a simple interpretation for the calculated
$\e_r<0$: Eqs.~\eqref{eq:er} correspond to Maxwell counting on the
$(d+1)$-core. While Maxwell argument is only approximate for the
complete graph, we propose that it is exact on the $(d+1)$-core, with
the onset of stressed nodes coinciding with the disappearance of
floppy modes; in other words, $\e_r$ gives the number of floppy modes
on the $(d+1)$-core when negative, and the number of redundant
constraints when positive. We verified numerically this interpretation
using the Pebble Game algorithm~\cite{JacobsThorpe95}, see
Fig.~\ref{fig:er}.

It follows that the rigidity percolation on the complete graph occurs
not at the $(d+1)$-core percolation threshold $\alpha_p^{(d)}$ but at
the point $\alpha_c^{(d)}$ where $\e_r$ becomes positive (for $d=1$,
the two thresholds coincide). The same conclusion was reached in
earlier studies~\cite{ThorpeJacobs99,Moukarzel03}, based on comparison
with numerical simulations.

%%%%%%%%%%%%%%%%%%%%%%%%%%%%%%%%%%%%%%%%%%%%%%%%%%%%%%%%%%%%%%%%%%%%%%%%%%

\begin{figure}
\centering\epsfig{file=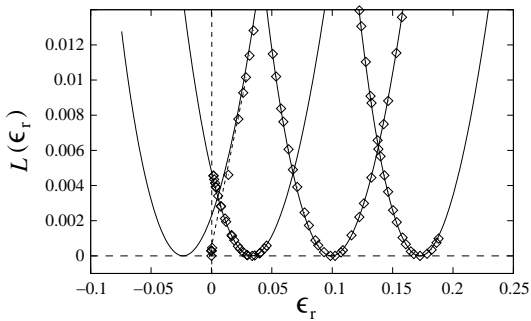,width=7cm}
\caption{Rate functions $L_{\rm cav}(\e_r)$ with $d=2$ and
$\alpha=1.75$, 1.85, 1.95 and 2.05 (from left to right at the bottom),
as obtained from Eqs.~\eqref{eq:cav}. The symbols are results of Monte
Carlo simulations at fixed $y$. For $\alpha<\alpha_{\rm Mxl}^{(2)}]$, the branches with $\e_r<0$ must be replaced by $L(\e_r<0)=\infty$. For
$\alpha=1.75<\alpha_c^{(2)}$, a part of the curve with $\e_r>0$ must
also be substituted by a Maxwell construction (dotted line).
\label{fig:ratefct}}
\end{figure}

\paragraph*{Finite temperature ---}

To analyze the consequences of allowing the network of constraints to
adapt, we now consider the system whose configuration space is the
ensemble $\G^{(\alpha)}_N$ of graphs with $N$ nodes and $M=\alpha N$
links. To each graph $G\in\G^{(\alpha)}_N$, we associate as energy the
number $E_r[G]$ of redundant constraints, as given by
Eq.~\eqref{eq:Er}. The partition function is thus $\Z_N(y)=\sum_{G\in
\G^{(\alpha)}_N}e^{-y E_r[G]}$ where the inverse temperature $y$
allows to sample different subsets of $\G^{(\alpha)}_N$, with the
infinite temperature limit $y=0$ corresponding to selecting the graphs
irrespectively of their energy, as considered previously. $\Z_N(y)$
can be evaluated by computing the microcanonical entropy density
$s(\e_r)$ which gives through $\exp [Ns(\e_r)]$ the number of graphs
in $\G^{(\alpha)}_N$ with density of redundant constraints $\e_r$. An
equivalent information is contained in the rate function $L(\e_r)$,
defined from the probability $\P_N(\e_r)$ for a graph in
$\G^{(\alpha)}_N$ to have density of constraints $\e_r$; indeed, if $\P_N(\e_r)$
satisfies a large deviation principle, $\P_N(\e_r)\asymp \exp [-N
L(\e_r)]$ ($a_N\asymp b_N$ means that $\ln a_N/\ln b_N\to 1$ as
$N\to\infty$), and, if $|\G^{(\alpha)}_N|$ denotes the number of
graphs in $\G^{(\alpha)}_N$, we have $\exp
[Ns(\e_r)]=|\G^{(\alpha)}_N| \exp [-NL(\e_r)]$. The equilibrium
properties of the system are thus captured by the potential $\phi(y)$,
with
\begin{equation}
e^{-N\phi(y)}=|\G^{(\alpha)}_N|^{-1}\Z_N(y)\asymp \int {\rm d}\e_r\
e^{N[y\e_r-L(\e_r)]},
\end{equation}
from which by Legendre transform we get
\begin{equation}
\e_r=\partial\phi(y)/\partial y,\qquad L(\e_r)=-y\e_r+\phi(y).
\end{equation}
This potential can be computed by the large deviation cavity
method~\cite{Rivoire05}, an extension of the cavity method to atypical
graphs. For random bond models, it yields:
\begin{equation}\label{eq:cav}
\begin{split}
&\phi(y)=-\ln Z+2\alpha\left[1-(1-\eta)e^{y}e^{-z}\right]-\alpha z,\\
&Z=e^{2\alpha\eta e^{-z}}+\sum_{r=0}^{d-1}\pi_{2\alpha\eta
e^{-z}}(k)\left(e^{-y(d-k)}-1\right),\\
&z=\ln\left[e^y+(1-e^y)\eta^2\right],\\
&\eta=1-\frac{1}{Z}\sum_{k=0}^{d-1}\pi_{2\alpha\eta
e^{-z}}(k)e^{-y(d-k)}.
\end{split}
\end{equation}
Examples of rate functions $L_{\rm cav}(\e_r)$ obtained from the
non-trivial solution ($\eta>0$) of these equations are displayed in
Fig.~\ref{fig:ratefct} for $d=2$ and four representative values of
$\alpha$; they are perfectly consistent with numerical results. For
$\alpha<\alpha_{\rm Mxl}^{(2)}=2$, $L_{\rm cav}(\e_r)$ extends to
$\e_r<0$. As we checked with numerical simulations, these unexpected
negative values have the same interpretation as in the typical case
($y=0$). This suggests that the correct rate function $L(\e_r)$ is
obtained from $L_{\rm cav}(\e_r)$ by truncating the negative branch:
$L(\e_r)=L_{\rm cav}(\e_r)$ if $\e_r>0$, and $L(\e_r)=\infty$ if
$\e_r<0$. For $d=1$, where $L_{\rm cav}(\e_r)$ never shows a negative
branch, our formul\ae\ coincide with rigorous results from the
mathematical literature~\cite{Puhalskii05} (we are not aware of any
previous result for $d\geq 2$).

%%%%%%%%%%%%%%%%%%%%%%%%%%%%%%%%%%%%%%%%%%%%%%%%%%%%%%%%%%%%%%%%%%%%%%%%%%

\paragraph*{Intermediate phase ---}

\begin{figure}
\centering\epsfig{file=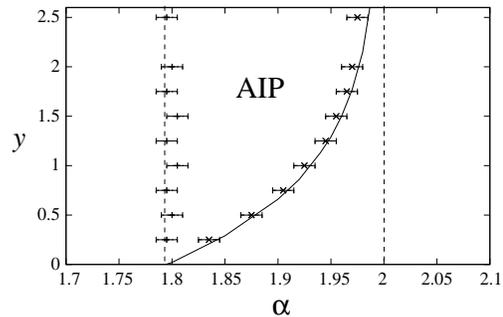,width=6.5cm} 
\caption{Phase diagram for $d=2$. The two vertical
dashed lines are for $\alpha_c^{(2)}=\alpha_c^{(2)}(0)$ and
$\alpha^{(2)}_{\rm Mxl}=2$, and the full line is for
$\alpha_c^{(2)}(y)$. At any fixed $y>0$, an intermediate phase is
present between $\alpha_c^{(2)}(0)$ and
$\alpha_c^{(2)}(y)$. The symbols are numerical estimates
for the locations of the two phase transitions.\label{fig:phasediag}}
\end{figure}

The rate function $L(\e_r)$ completely specifies the equilibrium
properties of the system for any $y$. Geometrically, the relation
$y=-\partial L(\e_r)/\partial \e_r$ indicates that $\e_r(y)$ is the
intersection of $L(\e_r)$ with the supporting line having slope $-y$
(the line below $L(\e_r)$ touching $L(\e_r)$ in a single point). For
$\alpha<\alpha_c^{(d)}$, we thus get $\e_r(y)=0$. For $\alpha>\alpha_c^{(d)}$,
the rate function has a decreasing part whose slope, by convexity, is
extremal at the lower edge, $\e_r=0$, with value denoted
$y_c^{(d)}(\alpha)=-\partial L/\partial \e_r (\e_r=0^+)$. If
$y<y_c^{(d)}(\alpha)$, which is necessarily the case for $d=1$ where
$y_c^{(1)}(\alpha)=\infty$, the density of redundant constraints is
$\e_r(y)>0$ and the system is in an UNSAT phase. For $d\geq 2$
however, since $y_c^{(d)}$ is finite when
$\alpha_c^{(d)}<\alpha<\alpha_{\rm Mxl}^{(d)}$, there is an
adaptive intermediate phase (AIP) between $\alpha_c^{(d)}$ where $y_c^{(d)}(\alpha)=0$ and
$\alpha_c^{(d)}(y)$ where $y_c^{(d)}(\alpha)=y$. In this phase, the
system maintains itself at the edge of the SAT-UNSAT transition,
with $\e_r(y)=0$. The resulting phase diagram is
presented in Fig.~\ref{fig:phasediag} for $d=2$.

\begin{figure}
\centering\epsfig{file=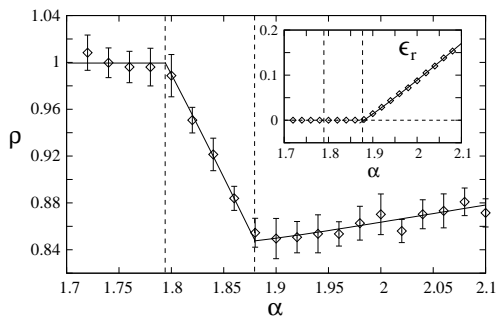,width=6.5cm} 
\caption{Order parameter $\rho=\sigma^2/(2\alpha)$ for $d=2$ when
varying $\alpha$ at fixed $y=0.5$. From the entry in the AIP at
$\alpha_c^{(2)}=1.79$, the networks are atypical (non-Poisson) random
graphs with $\rho<1$.  The symbols with error bars are from Monte
Carlo simulations. Inset: analytical and numerical results for
$\e_r(y)$.
\label{fig:oderpara}}
\end{figure}

The atypical, self-organized, nature of the graphs in the AIP can be
quantified with $\rho=\sigma^2/(2\alpha)$, the ratio of the variance
$\sigma^2$ of the degree distribution, over its mean
$2\alpha$. Indeed, typical random graphs have a Poisson degree
distribution~\cite{Bollobas01}, for which $\rho=1$, and a value
$\rho\neq 1$ is therefore indicative of atypical graphs. We computed
$\rho$ with the large deviation cavity method, and found, as shown in
Fig.~\ref{fig:oderpara}, that it presents two slope discontinuities,
at the two boundaries of the AIP; the density of redundant
constraints $\e_r$ in contrast is strictly zero until $\alpha$ reaches
$\alpha_c^{(d)}(y)$. This result is in very good agreement with our
numerical simulations, and clearly demonstrates that the transition at
$\alpha_c^{(d)}$ is of a different, topological, nature than the
SAT-UNSAT transitions taking place here at
$\alpha_c^{(d)}(y)>\alpha_c^{(d)}\equiv \alpha_c^{(d)}(y=0)$.

%%%%%%%%%%%%%%%%%%%%%%%%%%%%%%%%%%%%%%%%%%%%%%%%%%%%%%%%%%%%%%%%%%%%%%%%%%

\paragraph*{Zero temperature ---}

It is worth mentioning that the extreme case $y=\infty$, which has
been the focus of several numerical
studies~\cite{ThorpeJacobs00,ChubynskyBriere06} appears here as a very
singular limit. In this case indeed, the distinction made previously
between $y_c^{(d)}$ being infinite or not is irrelevant since $y$
itself is infinite. An asymptotic analysis of our solution reveals
that when $y=\infty$ an AIP is present between $\alpha_c^{(d)}$ and
$\alpha_{\rm Mxl}^{(d)}$ for any value of $d$, including $d=1$ for
which it disappears as soon as $y<\infty$. This suggests that in
finite dimensional models, where the rigidity transition is thought to
be continuous, the AIP could be confined to $y=\infty$. An AIP at
finite $y$ may be present only as cross-over, which could also account
for the experimental observations in network glasses.

%%%%%%%%%%%%%%%%%%%%%%%%%%%%%%%%%%%%%%%%%%%%%%%%%%%%%%%%%%%%%%%%%%%%%%%%%%

\paragraph*{Discussion ---}

CSP's such as $K$-SAT or coloring are known for displaying a clustered
intermediate phase (CIP), also known as hard-SAT phase, where the set
of solutions breaks into an exponential number of disconnected
components~\cite{MezardZecchina02,MuletPagnani02}. This CIP, which, in
contrast to the AIP, takes place prior to the UNSAT-SAT transition, is
however of a radically different character: it refers to a transition
in the space of solutions associated to single, typical, graphs, and
not to a transition in a graph space. We analyzed here random bond
models, which we argued do not involve replica symmetry breaking, and
thus have no CIP. CSP's with a CIP could however be tackled along the
same lines to study the interplay between the two kinds of
intermediate phases.

In conclusion, we presented a general analytical approach for solving
the equilibrium thermodynamics of models of adaptive networks. This
leads us to the exact description of a new kind of topological phase
transition, which takes place in a configuration space made of
graphs. Our findings confirm that the presence of a discontinuous
SAT-UNSAT transition in the typical problem is essential for observing
an adaptive intermediate phase (AIP)~\cite{BarreBishop05b}.  The
principles behind the self-organization of graphs found in the AIP
differ notably from most of the mechanisms proposed so far to explain
the non-random structure of networks. Our model indeed belongs to the family of
exponential random graphs, a still poorly understood class of
networks~\cite{Newman03}. Our results show that the cavity method
for large deviations is a powerful tool to study at least some of
these models.

\paragraph*{Acknowledgments ---} We warmly thank Mykyta Chubynsky, for 
making available to us his implementation of the Pebble Game
algorithm. O.R. is a fellow of the Human Frontier Science Program.

\bibliographystyle{apsrev}

\bibliography{/home/Oliv/REF/physref}

\end{document}